\documentclass{article}
\usepackage{scalerel}
\usepackage{adjustbox}
\usepackage{blindtext, rotating}
\usepackage{amsthm}
\usepackage{bbm}
\usepackage{amsmath}
\usepackage{amssymb}
\usepackage{natbib}
\usepackage{enumerate}
\usepackage{graphicx}
\usepackage{amssymb}
\usepackage{multirow}
\usepackage{verbatim}
\usepackage{subfigure,placeins}
\usepackage{array}
\newcolumntype{C}[1]{>{\centering\arraybackslash}p{#1}}

\usepackage{enumerate}
\usepackage{bbold}

\usepackage{latexsym}
\usepackage{amsmath}
\usepackage{amsfonts}
\usepackage{amssymb}
\usepackage{psfrag}
\usepackage{graphicx}
\usepackage[dvipsnames]{xcolor}
\usepackage{url}
\usepackage{bm}
\usepackage[ruled,vlined]{algorithm2e}
\usepackage{mathtools}
\usepackage{multirow}
\usepackage{dcolumn} 
\newcolumntype{d}[1]{D..{#1}}
\usepackage{verbatim}
\usepackage{soul}
\usepackage{tikz}
\usetikzlibrary{shapes,decorations,arrows,calc,arrows.meta,fit,positioning}

\title{Causal inference: critical developments, past and future}
\author{Erica E. M. Moodie$^1$ \and David A. Stephens$^2$ \\ \ \\
	$^1$Department of Epidemiology and Biostatistics, McGill University, Canada\\
	$^2$Department of Mathematics and Statistics, McGill University, Canada\\
	}

\date{\today}
	
\begin{document}
		
\maketitle

		\begin{abstract}
			Causality is a subject of philosophical debate and a central scientific issue with a long history.  In the statistical domain, the study of cause and effect based on the notion of `fairness' in comparisons dates back several hundred years, and yet statistical concepts and developments that form the area of causal inference are only decades old. In this paper, we review core tenets and methods of causal inference and key developments in the history of the field. We highlight connections with traditional `associational' statistical methods, including estimating equations and semiparametric theory, and point to current topics of active research in this crucial area of our field. 

		\end{abstract}

	\section{Introduction}

In this paper, we review the rapidly developing literature on causal inference from a statistical perspective.  Causal thinking has long guided study design and analysis, but only relatively recently have the core ideas come into the statistical mainstream.  Alongside the development of paradigms for expressing priorities in a causal setting, statistical frameworks and methods have been the focus of much research.

A central objective in many statistical settings is to infer (or assess, or test for) the impact of an intervention on an outcome on the basis of study data.  The central challenge is that, typically, the data available are not suitable for making the assessment without further assumptions or adjustment, as, although the intervention can be observed in the data, it is done so in an `uncontrolled' fashion, and it is plausible that any observed effect may be due to other factors.  Fundamentally, it is this `confounding' of the intervention effect that must be accounted for using statistical techniques.

In modern statistical research, and in related areas in machine learning, there is considerable mystique and many misconceptions regarding the nature of causal inference. As we will discuss, it is false to suppose that any analysis simply using, say, a model for the treatment allocation (a `propensity score') will return a causal estimate. \textit{At the same time,}  it is equally false to suggest that ordinary, conditional regression models \textit{cannot} provide causal estimates: although there are some settings in which this is true, it is not universally the case.

Some of the most prominent and important leaders in the field of causal inference have, unfortunately, contributed to these views. For example, in an introductory paper on causal inference, \citet{Pearl2010} writes `` causal and associational concepts do not mix'' and follows this by stating
\begin{quote}
A useful demarcation line between associational and causal concepts crisp and easy to apply, can be formulated as follows. An associational concept is any relationship that can be defined in terms of a joint distribution of observed variables, and a causal concept is any relationship that cannot be defined from the distribution alone.
\end{quote}
Whereas this point is well-taken from a philosophical perspective, in statistical terms it creates a false dichotomy, and one that regrettably often serves to create tension where none need exist. Traditional, non-predictive statistics typically do aim to answer causal questions. The core tenet of causal inference is to do so through careful consideration of the context and through \textit{explicit} statements about the assumptions regarding the underlying data-generating mechanism (`nature') as well as the estimands (what contrasts, applied to which population).

\section{Core ideas} \label{sec:core}
Causal inference typically relies on several identification assumptions which correspond to implicit assumptions made in traditional regression analyses, as we shall detail below. The formulation of causal inference problems, however, begins several steps earlier, with overt focus on \textit{estimands}, or target parameters of interest, which are defined in terms of contrasts between treatment levels within a specific population.

A causal approach to an analysis must first define the causal research question, and then typically proceeds as follows \citep{STRATOSpaper}:
\begin{enumerate}[A.]
\item Formalize all definitions:
\begin{enumerate}[1.]
\item define the treatment and its relevant levels,
\item define the outcome (specifying, if necessary, the time following treatment at which it is measured),
\item define the population(s) of interest.
\end{enumerate}
\item Specify the estimand (target causal effect).
\item Perform the estimation:
\begin{enumerate}[1.]
\item state the identifiability and estimation assumptions,
\item estimate the causal effect,
\item evaluate assumptions and perform sensitivity analyses.
\end{enumerate}
\item Perform inference.
\end{enumerate}
Causal estimands (step 2) are often formulated in terms of potential outcomes, also known as counterfactuals. A critical component to determining the assumptions (step 3.1) requires the analyst to determine via existing literature or substantive experts the relevant variables needed to posit a realistic model for the data generating structure. This may be formalized in a directed acyclic graph (DAG), as shall be detailed below. Estimating the causal effect can be often be accomplished using a variety of approaches, which include both traditional statistical methods and those considered to be ``causal approaches" such as propensity score based methods. Some variations on the above road-map have been proposed \citep[e.g.][]{Toolkit2019}, however the focus on formalizing and making explicit all assumptions is common.

The final step, of performing inference, is afflicted by the same challenges in the causal framework as in standard statistical settings. Calculating measures of variability or (un)certainty such as standard errors or confidence intervals is often computationally straightforward. As we shall see in Section \ref{sec:estimate}, many causal methods of estimation require two-step estimators, relying on plug-in estimates and/or requiring marginalization of conditional models. Asymptotically, many such methods do not require correction for the substitution estimator \citep{HenmiEguchi2004}; nonetheless, a standard, non-parametric bootstrap is commonly employed to guard against optimism in variance calculations that ignore nuisance parameter models in finite samples. There, the simplicity ends, as decision-making -- drawing conclusions about ``significance" of a finding -- is no less controversial in the causal realm as in standard (frequentist) statistics \citep{GreenlandUncondPval2019}.


As above, we denote intervention or treatment by $Z$, pre-treatment covariates by $X$, and observed outcomes by $Y$. We denote potential outcomes, defined in the following section, by $Y(z)$ for all $z$ in some treatment set $\mathcal{Z}$; loosely, we can consider the potential outcome $Y(z)$ to be the random variable that records the outcome that would occur (perhaps contrary to observation) if the treatment was set to $z$ by intervention of the researcher.  In, say, agricultural and medical settings, the concept of a controlled experiment, where intervention occurs by design, is long-established, and the natural setting within which to initiate the consideration of causal concepts.  We expand on this in the following section.

\section{Key historical developments}
\subsection{Randomization and confounding}
Though perhaps only really recognized in hindsight \citep{RubinDiscussion1990}, Neyman is now credited with being the first to describe the notion of a potential outcome \citep{Neyman1923}, when he described the (unknown) ``potential'' yield of plots under varying conditions. This concept of potential outcomes has been used to formulate an entire framework \citep{Rubin1974} on which much of modern causal inference is based, however notions of causality in statistics, particularly medical statistics, arose earlier.

The first known \textit{clinical} trial is often attributed to James Lind's 1747 study of citrus to treat scurvy \citep{LindTrial}.
The first randomized clinical trial occurred much later \citep{MRC1948} following principles laid out by Sir Austin \citet{ABH-book1937} in the previous decade, which built on the design of experiments of Sir Ronald Aylmer Fisher from the 1920s and beyond; see e.g.~\citet{FisherDesign1951}, or, for an excellent summary, \cite{BoxFisher}. While these works are foundational and formalize statistical issues around the notion of confounding, the idea was not entirely new: recent work \citep{Chalmers2011} has highlighted notions of `fairness' which date back to the fourteenth century \citep{Petrarca1364}:
\begin{quote}
I solemnly affirm and believe, if a hundred or a thousand men of the same age, same temperament and habits, together with the same surroundings, were attacked at the same time by the same disease, that if one half followed the prescriptions of the doctors of the variety of those practising at the present day, and that the other half took no medicine but relied on Nature's instincts, I have no doubt as to which half would escape.
\end{quote}
It is this concept of `fairness' in comparison groups, or lack of confounding, that is central to causal comparisons. Randomization aims to find groups of individuals whose covariates (e.g., the age, temperament, and habits of Petrarca) follow the same distribution and allow only one factor -- the treatment of interest -- to vary systematically. Potential outcomes provide a notation and a means of expressing this at an individual level. That is, for a binary treatment $Z \in \{0,1\}$, we may consider the outcomes of an individual -- whose covariates are fixed and unchanging -- under the two values of treatment, and we label these random variables $Y(0), Y(1)$, respectively. Potential outcomes with individual index $i$ have variably been denoted $Y_z(i)$, $Y_i(z)$, $Y_i^z$, or by distributional equivalents such as $P(Y = y|\text{set}(Z = z))$ or $P(Y = y|\text{do}(Z = z))$, where the latter form the basis of what is known as Pearl's ``do-calculus'' (\citeyear{Pearl_Causality})
and can be used to consider outcome distributions or estimands based on, say, expectations of these distributions at the population level. Extensions of potential outcomes to multi-valued or continuous treatments are straightforward.

\subsection{Towards drawing causal conclusions}
Prior to the formalization of causal inference via the potential outcomes framework, however, rather less rigourous but important discussions of causality arose in the field of medical statistics. Perhaps the best known ideas are what are often referred to as the ``Bradford Hill criteria'' proposed in `The Environment and Disease: Association or Causation?' \citep{ABH-criteria}:
strength of relationship; consistency; specificity; temporality (cause precedes effect); biological gradient; plausibility; coherence; experiment (trials); and analogy.

In fact, Bradford Hill specifically noted that these were \textit{not} criteria, but rather a group of conditions that might be useful to \textit{assess} (rather than establish) causality.  In an earlier lecture, \citet{ABH-MedStat} asked
\begin{quote}
The question, on the other hand, may well be asked, what does one accept as overwhelming?
\end{quote}
\noindent and later stated
\begin{quote}
We are continuously brought back to the fundamental question -- what alternative explanation will fit a set of observations, what other differences between our contrasted groups could equally, or better, account for the observed incidences.
\end{quote}
\noindent The answer lies, Bradford Hill suggests, at least in part in a thorough understanding of the context and the science of the substantive question to be answered, and ``is not to be found tidily tucked up in the formulae of tests of significance, useful as they may be."

Bradford Hill's 1965 article was reprinted in 2020 of the journal \textit{Observational Studies}, and the 1962 lecture reprinted in \textit{Statistics in Medicine} not long after, in each case followed by commentary and discussion \citep{ABH-criteria-ObsStudy,ABH-MedStat-StatMed}. These reflections highlight the many ways in which Bradford Hill was a visionary statistician and are a call to reflect on the manners in which current data and research questions may require different or additional assumptions and considerations.

\subsection{Designing observational studies to permit causal conclusions} \label{sebsec:Rubin}
In a seminal paper, \citet{Rubin1974} described a variety of strategies that could be used to estimate causal effects from observational (non-randomized) studies. This article was a rebuttal to the then prevailing wisdom that causal effects could only be learned from randomized studies. As Rubin noted, there are many instances where randomization is infeasible (e.g., due to cost or a long latency period of the exposure) or unethical (e.g., the exposure of interest is harmful). \citet{Rubin1974} began by defining an average causal (or treatment) effect (ATE) as  \[ E[Y_i(1) - Y_i(0)],\] where, as above, $Y_i(z)$ for $z = 0$ or $1$ denotes the potential outcome for individual $i$ if they were to receive treatment $z$. Noting that these two potential outcomes can never simultaneous be observed, he remarked that $Y_j(z)$ may serve as a good stand-in for $Y_i(z)$ if individuals $i$ and $j$ are ``perfectly matched''. If the entire sample consisted of $n$ such matched pairs, then the difference in sample average among those treated and those not can estimate the ATE, and even if matches are imperfect, the result should be close. Drawing an analogy between the randomized and observational studies, Rubin noted that chance imbalance can occur even under randomization, and ``we cannot be guided solely by the concept of unbiasedness over the randomization set. We need some model for the effect of prior variables in order to use their values in an intelligent manner." A concurrent publication \citep{CochranRubin1974} points to regression and matching as two such means of addressing bias on observational studies.

Over the next several years, Rubin (e.g.,~\citeyear{rubin_assignment_1977}, \citeyear{Rubin1979}) produced a series of papers focusing, primarily, on matching and regression to remove bias in estimation due to measured confounding variables in observational studies and laid the foundations for the ground-breaking development of the propensity score \citep{RosenbaumRubin1983}. The propensity score is the coarsest of all balancing scores, where a balancing score is a function of (pre-treatment) covariates such that, conditional upon it, treatment and those covariates are independent. More formally, a balancing score $b(x)$ is any function of covariates $x$ such that $Z \perp X | b(x)$. Trivially, $x$ is a balancing score. Note that for a balancing score to be of any use, the covariates $X$ on which balance is produced should include all relevant confounders (or at least the subset of all confounders needed to ensure that no spurious association between $Z$ and $Y$ may be found; see Section \ref{sec:DAGs} for further discussion).

Denote by $e(x)$ the propensity score, defined by $e(x) = \Pr(Z=1|x)$. It transpires \citep{RosenbaumRubin1983} that $Z \perp X | e(x)$ and further that $(Y(0),Y(1)) \perp Z | e(x)$, implying strong ignorability given the propensity score. This result may be used to establish that the ATE for a continuous outcome $Y$ can be found via (i) constructing a matched sample using the (univariate) propensity score to perform matching rather than the typically higher-dimensional covariate vector, (ii) subclassification or stratification based on the propensity score, or (iii) a linear regression model in which the balancing score is included as a covariate, in what is known as `propensity score regression', where $e(x)$ is used as a predictor in the regression model.  Note that this final approach encompasses -- under certain assumptions -- traditional regression analysis since the full vector $x$ is a balancing score; see Section \ref{sec:reg}.  \citet{RosenbaumRubin1983} further demonstrated that an estimated propensity score retains the desired balancing properties in finite samples provided the estimated propensity score is bounded away from 0 and 1.

In parallel to the literature on adjustment for confounding, a growing literature on missing data emerged, partly reflecting the fact that the potential outcomes themselves are largely `missing'. \citet{RRZ1995-IPW} proposed semiparametric estimators known as `inverse probability weighted estimators' for missing data, where the probability used to weight was that of an observation being recorded conditional on measured covariates. Under the setting in which data are missing at random \citep{Rubin1976-MissingBka} and missingness probabilities are bounded away from 0 and 1, the derived estimators are asymptotically consistent and semiparametric efficient. From this point, the connection to confounding was drawn: as \citet{Holland1986} had previously noted, the `fundamental problem' in causal inference is the impossibility of observing both $Y(0)$ and $Y(1)$ for any individual. This provides a missing data lens through which to view potential outcomes, and led to the development of inverse probability of treatment weighting as a means of addressing confounding to estimate parameters from marginal models \citep{Robins1998msm}.

Through the 1990s and beyond, both methods of analysis and, importantly, \textit{designs} of observational studies continued to be developed, with the objective reducing or eliminating bias due to confounding. For example, \citet{Rubin2007Design} proposed the construction of an analytic dataset in an observational context such that treatment groups have balanced (comparable) covariate distributions without any knowledge of the outcome variable, via an approach that aims to emulate a randomized trials using techniques such as matching. These ideas continue to be repeated and refined to meet the needs of ever-growing data sources, many of which are not only observational, but also not collected for research purposes \citep[e.g.][]{HernanRobinsTargetTrial,DanaeiEtAl2018,Murray2021emulating}.


\subsection{When tradition fails (almost)} \label{sebsec:msm}
As we will demonstrate in Section \ref{sec:estimate}, there are many estimands for which traditional statistical approaches such as regression adjustment can - under assumptions of correct model specification - yield consistent estimators. This includes the average treatment effect, the causal risk ratio, and so on. However in the last few decades, the causal literature has produced some estimands that are difficult  to define without the aid of potential outcomes, even though it is possible to estimate these quantities using some combination or repeated application of traditional methods such as regression, prediction, and averaging.

Consider, for example, the average treatment effect on the treated (ATT): $E[Y_i(1) - Y_i(0)|Z=1]$ \citep{ImbensAngrist1994-LATE}. This estimand captures the effect of treatment in the population who were in fact treated, a population whose covariates typically differ from the population as a whole and for whom the treatment effect may be most relevant. Consider, for example, a nicotine substitution treatment. The impact of this treatment on number of cigarettes smoked is not relevant to the whole population or perhaps even on the population of smokers, but the causal effect of the nicotine substitution among those who are willing to accept treatment -- and therefore are showing an interest or commitment in giving up smoking -- is arguably the most relevant target population. While it is possible to estimate this quantity using regression and prediction, it is more commonly and more easily estimated using propensity score methods and arguably an estimand that is less easily formulated without `causal thinking': potential outcomes are not necessary to define the ATT, but prove useful since the expectation $E[Y_i(0)|Z=1]$ is not commonly considered in probabilistic modelling.

Another instance in which potential outcomes proved useful for the formulation of an important and meaningful outcome came with the introduction of marginal structural models \citep{Robins1999,Robins1999AssCauseMSM,RobinsEtAl2000}. Marginal structural models are typically defined in a setting where there is a longitudinal sequence of treatments - say $Z_1$ and $Z_2$, and interest lies in the impact of that sequence of treatments on some final outcome $Y$, such that $E[Y_i(z_1,z_2)]$ is of interest, or a contrast between this expectation for a given pair of treatment combinations, e.g.~$E[Y_i(1,1)-Y_i(0,0)]$. Note that this model is \textit{marginal} rather than conditional, in that it marginalizes over pre-treatment covariates, including any intermediate variables that might be affected by $Z_1$ but precede $Z_2$. (The term \textit{structural} is used to indicate that the models is assumed to be causal, with estimands encompassing effects that can be derived from the true data-generating relationships.)  It is now well-known that such situations where an intermediate variable is both a mediator of the relationship between $Z_1$ and $Y$ and a confounder of the $Z_2$ and $Y$ relationship, there is no regression model parameter that corresponds to $E[Y_i(z_1,z_2)]$ \citep[e.g.][]{HernanEtAl2000,Moodie2010_IJPH}: a model that omits the intermediate variable will not appropriately adjust for confounding of the effect of $Z_2$ and yet including that variable in the model will bias the estimator of the effect of $Z_1$ by blocking any part of its impact that is mediated through that variable.

As an aside, we note that while marginal structural models are most often used in the context of time-varying exposures, these models can also be used in a point-treatment setting. In fact, the average treatment effect ATE = $E[Y(1) - Y(0)]$ is a marginal structural model that marginalizes over all pre-treatment covariates. Marginal models may also be only `partially' marginal, for example it is possible to estimate a treatment effect that conditions on some particular covariate (older age, say, or a specific co-morbidity) and marginalizes over all other covariates.

In this section, we have provided a partial and certainly brief survey of developments in statistical causal inference. We now turn to a formalization of the statistical assumptions as well as some typical methods of estimation employed to estimate quantities such as the average treatment effect.

\section{Assumptions and estimation}

\label{sec:Assess}
In this section, we formalize the tools and assumptions made in causal inference, and provide a more technical description of the estimation approaches outlined in Sections \ref{sebsec:Rubin}-\ref{sebsec:msm}.

\subsection{Causal graphs} \label{sec:DAGs}
Directed Acyclic Graphs (DAGs), often called causal graphs, are a tool used to visually display causal assumptions that are made about the data generating relationships under consideration in a given substantive problem, and to assist in determining those variables that can distort or bias the causal effect(s) of interest \citep{Pearl995}. A graph is said to be directed if all inter-variable relationships are connected by arrows, where an arrow from one variable into another indicates that the first variable causes changes in the second. The graph is said to be acyclic if it has no closed loops. Finally, for a DAG to be causal, all common causes of the treatment $Z$ and outcome $Y$ (measured or otherwise) must be included in the graph \citep{VanderWeeleDAGs2008}.

Consider, for example, the left panel of Figure \ref{fig:DAGs}: in this very simple DAG, we see that $X$ is a common cause (and therefore a confounder of the relationship between) $Z$ and $Y$, while $M$ is a mediator in that relationship. DAGs can be elaborated to include more covariates, e.g., by considering confounders separately or including variables that may affect $Y$ but not $Z$, or can be used to depict treatment sequences, with treatments and intervening variables observed an multiple time-points as in the right panel of Figure \ref{fig:DAGs}. The aim of causal inference is to quantify the `flow' of the effect of an intervention on treatment $Z$ through all the directed paths that connect $Z$ to $Y$, while accounting for all the other paths that connect the two variables.  For example, in the left panel of Figure \ref{fig:DAGs}, there are two paths by which $Z$ may act on $Y$ (one direct $Z \rightarrow Y$, the other passing through $M$); in contrast, there is an `open' path $Z \leftarrow X \rightarrow Y$ that is undirected, and via which  a co-variation of $Z$ and $Y$ may be observed unless the path is blocked, which is typically achieved by conditioning on $X$.  The path $Z \leftarrow X \rightarrow Y$ is termed a confounding or backdoor path, and conditioning on $X$ removes the confounding by blocking this path. For an excellent introduction to the use of causal DAGs to identify those variables needed to block any spurious paths between $Z$ and $Y$, we refer the interested reader to \citet{GreenlandEtAl1999} or \citet{DidelezChapter2018}. In the right panel of Figure \ref{fig:DAGs}, the situation is more complex, and conditioning strategies to remove confounding are less obvious \citep{Moodie2010_IJPH}.

Figure \ref{fig:DAGPS} illustrates the role of the propensity score in adjustment; the propensity score $e(X)$ is a deterministic function of the confounders $X$, and knowledge of $e(X)$ determines the distribution of $Z$, and renders $Z$ conditionally independent of $X$. Recall that $e(X)$ is univariate irrespective of the dimension of $X$.

Causal graphs are useful tools for displaying and clarifying the proposed relationships between variables, and for formulating causal questions. However, whereas DAGs encode specific probabilistic relationships, they do not uniquely define joint probability distributions; this re-affirms the comment of \cite{Pearl2010} mentioned in the introduction.  Consequently, learning causal structure from observed data is a very challenging problem, which re-inforces the importance of the strategies outlined in Section \ref{sec:core}, and Bradford Hill's pragmatic problem formulation approach.

\tikzset{
    -Latex,auto,node distance =1 cm and 1 cm,semithick,
    state/.style ={draw, minimum width = 0.7 cm},
    box/.style ={rectangle, draw, minimum width = 0.7 cm, fill=lightgray},
    point/.style = {circle, draw, inner sep=0.08cm,fill,node contents={}},
    bidirected/.style={Latex-Latex,dashed},
    el/.style = {inner sep=3pt, align=left, sloped}
}
\begin{figure}
\centering
\begin{tikzpicture}[scale=1.75]
    \node (x1) at (-1.5,1.75) {${X}$};
    \node (y1) at (1,1){${Y}$};
    \node (z1) at (-1,1) {${Z}$};
    \node (m1) at (0,1) {${M}$};

    \path (x1) edge (z1);
    \path (x1) edge (y1);
    \path (z1) edge (m1);
    \path (m1) edge (y1);
    \path (z1) edge[bend right] (y1);


    \node (x21) at (2.5,1.75) {${X_1}$};
    \node (x22) at (3.5,1.75) {${X_2}$};
    \node (y2) at (5,1){${Y}$};
    \node (z21) at (3,1) {${Z_1}$};
    \node (z22) at (4,1) {${Z_2}$};
    \node (u2) at (4.5,2.25) {${U}$};

    \path (x21) edge (x22);
    \path (x21) edge (z21);
    \path (x22) edge (z22);
    \path (z21) edge (x22);
    \path (z21) edge (z22);
    \path (z22) edge (y2);
    \path (x22) edge (y2);
    \path (u2) edge (y2);
    \path (u2) edge (x21);
    \path (u2) edge (x22);
    \path (z21) edge[bend right] (y2);

 \end{tikzpicture}
    \caption{Examples of Directed Acyclic Graphs. The left panel shows a simple setting with a single treatment, $Z$, whose effect on the outcome $Y$ is confounded by $X$ and mediated through another variable, $M$. The right panel is a more complex, longitudinal example with two treatments, $Z_1$ and $Z_2$ that occur in sequence, each temporally preceded by covariates $X_1$ and $X_2$, and a final outcome $Y$. A possibly unobserved variable $U$ causally affects $X_1$, $X_2$, and $Y$. }
    \label{fig:DAGs}
\end{figure}
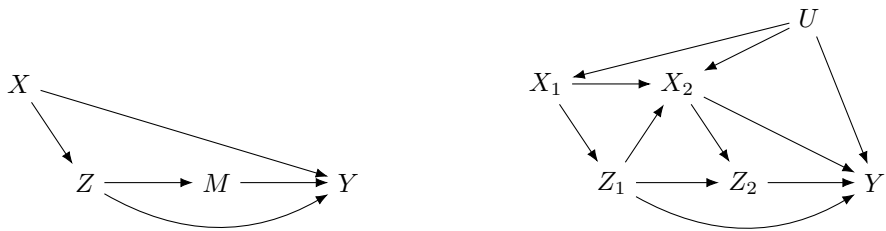

\tikzset{
    -Latex,auto,node distance =1 cm and 1 cm,semithick,
    state/.style ={draw, minimum width = 0.7 cm},
    box/.style ={rectangle, draw, minimum width = 0.7 cm, fill=lightgray},
    point/.style = {circle, draw, inner sep=0.08cm,fill,node contents={}},
    bidirected/.style={Latex-Latex,dashed},
    el/.style = {inner sep=3pt, align=left, sloped}
}
\begin{figure}
\centering
\begin{tikzpicture}[scale=1.75]
    \node (x1) at (-1.5,2.5) {${X}$};
    \node (e1) at (-1.25,1.75) {$e(X)$};
    \node (y1) at (1,1){${Y}$};
    \node (z1) at (-1,1) {${Z}$};

    \path (x1) edge (e1);
    \path (e1) edge (z1);
    \path (z1) edge (y1);
    \path (x1) edge (y1);

  \end{tikzpicture}
    \caption{The role of the propensity score: $e(X)$ sits on the path between $X$ and $Z$, and conditioning upon $e(X)$ renders $Z$ independent of $X$ by blocking the path from $X$ to $Z$.}
    \label{fig:DAGPS}
\end{figure}
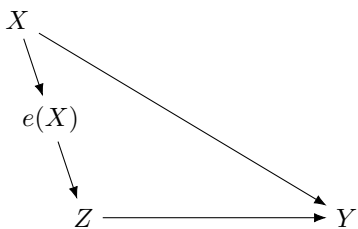

\subsection{Identifiability assumptions}
There are several key assumptions that are made in much of the causal literature. While at first these may appear to be restrictive, we shall see that in fact these assumptions are often made implicitly in the standard statistical inference literature.

\textit{Consistency}, also sometimes referred to as the well-defined treatment assumption or treatment variation irrelevance, states that any variations in the exposure of interest have no relevance to its impact on the outcome. This could be true in the setting where the treatment is a dietary supplement given either as a capsule or in drops, but is unlikely to hold when the `treatment' is ``5kg weight loss" which could be accomplished by exercise, caloric intake restriction, or some combination of these. While weight loss caused by both exercise and caloric intake may lead to reduced joint pain, the impact of the method of weight loss may differ on outcomes such as mood or overall self-reported well-being. This assumption is primarily made to ensure a `clean' and well-defined, reproducible result rather than for statistical purposes. This assumption is rarely if ever stated in traditional statistical inference literature, and yet few would disagree with its utility in terms of making causal statements. Under consistency, potential outcomes are linked to observed data via the equality $Y = Y(0)(1-Z) + Y(1)Z$.

A type of independence assumption commonly made in the causal literature is known as the \textit{Stable Unit Treatment Value Assumption} (SUTVA); this requires consistency and further states that an individual's outcome is affected only by the treatment they receive, but not that of others \citep{Cox1958,RubinJASA1980}. This is also referred to as a lack of `interference' or `spill-over', and is most easily explained in terms of examples where it does not hold. Consider, for example, the case of a treatment by vaccination: the vaccination status of the members of one's household (workplace, community, etc.) has significant bearing on an individual's probability of encountering and hence catching an illness such that the outcome of that individual is affected by the treatment status of other individuals. SUTVA is thus unlikely to hold when treatment is a vaccine.

The assumption of \textit{no unmeasured confounding} is perhaps the strongest and also the least (statistically) controversial. Stated in terms of potential outcomes, it requires $\{Y(0),Y(1)\} \perp Z | X$ -- that is, given the available information contained in covariates $X$, the treatment received carries no additional information on the individual's potential outcomes. This assumption is also referred to as conditional exchangeability or strong ignorability \citep{RosenbaumRubin1983}. In more traditional statistical terms, this is equivalent to saying that treatment is randomized within strata defined by $X$. Other forms of this assumption exist; for example, weak ignorability requires $Y(z) \perp Z | X$ for each potential outcome $Y(z)$ but not, as in strong ignorability, jointly for both potential outcomes \citep{GreenlandRobins2009}.

Finally, the assumption of (conditional) \textit{positivity} or overlap states that there exists no strata of $X$ such that treatment assignment is uniquely determined (or indeed, some levels of treatment are impossible): $P(Z=z|X=x) > 0 \ \forall z, x$. This assumption is often stated in the context of analyses that rely on the propensity score and is a necessary assumption for non-parametric estimation of the ATE. However, positivity can be viewed through the lens of extrapolation in a more classical estimation setting. Suppose that for a given value of $X$, say $x^*$, $P(Z=0|X=x^*) = 1$. It is clear that there will be no triples of observed data $(Y,Z,X)$ such that $Z=1$ and $X=x^*$; thus, any estimates of, say a conditional average treatment effect such as $E[Y|Z=1,X=x^*] - E[Y|Z=0,X=x^*]$ would require extrapolation. Of course, there may be instances where we are willing to do so, e.g., if we are willing to assume correct specification of the conditional mean model for $Y$. If such a model cannot be assumed to be correct (whether locally, over the region where all pairs $(z,x)$ are observed, or more globally), then positivity is required.

\subsection{Commonly employed approaches to estimation of causal effects} \label{sec:estimate}
Following a careful review of existing literature and consultation with subject matter experts, an analyst should identify a sufficient set of confounders $X$, possibly using a DAG. Assuming these confounders are all measured without error, the  causal effect of treatment can be estimated by comparing observed outcomes between different levels of treatment amongst groups of individuals with identical or very similar values of  $X$, where that comparison can be performed in a variety of ways including regression, stratification, matching, or weighting.

\subsubsection{Traditional outcome regression modelling}
\label{sec:reg}
Regression modeling can be used to undertake a causal analysis. A model must be specified for the outcome as a function of the treatment $Z$ and confounders $X$. For instance, one could suppose a linear model such as
\[ E[Y|Z,X,\beta] = \beta_0 + \beta_{Z}Z + \beta_{X}'X \]
when the outcome is continuous, such that the ordinary least squares estimator of $\beta_{Z}$ is a consistent estimator of the ATE provided the above model is correctly specified (i.e.,~the expected value of $Y$ depends linearly on $Z$ and $X$ and there are no interactions between these variables). Alternatively, a more complex model could be fit, allowing for, say, polynomial terms and treatment-covariate interactions. In this latter case, obtaining an estimator of the ATE requires an additional step of averaging predictions over the empirical distribution of the confounders $X$. That is, we posit a model such as
\begin{equation} \label{eqn:outcomeReg}
E[Y|Z,X;\beta] = \beta_{X,Z}'h(X, Z)
\end{equation}
where $\beta_{X,Z}$ is a vector of parameters and $h(X, Z)$ is a design matrix containing main effects of the treatment, the covariates, transformations of these which might include interactions and polynomial functions. Assuming this model is structural (captures the true data generating mechanism, i.e.~models the potential outcomes), then the ATE may be estimated from a sample of size $n$ as
\begin{eqnarray*}
 \widehat{E}[Y(1) - Y(0)] &=& n^{-1}\sum_{i=1}^n \widehat{Y}(1) - n^{-1}\sum_{i=1}^n \widehat{Y}(0) \\
      &=& n^{-1}\sum_{i=1}^n \hat\beta_{X,Z}'h(X_i, 1) - n^{-1}\sum_{i=1}^n \hat\beta_{X,Z}'h(X_i, 0) .
 \end{eqnarray*}
Note that if the estimand of interest was the ATT rather than the ATE, the marginalization (empirical averaging) would simply be taken over only those individuals observed to be treated in the sample, rather than the entire sample. Thus, the average would be taken over the distribution of $X|Z=1$. Note that there is no computational or conceptual barrier to computing a predicted outcome, $\hat\beta_{X,Z}'h(X_i, 0)$, under the condition $Z=0$ for those individuals for whom $Z$ is in fact observed to be 1.

\subsubsection{Propensity score regression}
\label{sec:PSreg}
Propensity score regression can be viewed as a special case of outcome regression modelling, however the assumptions required for consistency differ. Under this approach, again a model is specified as in equation \eqref{eqn:outcomeReg}, however in this specification, the design matrix $h(X, Z)$ includes as columns (`predictors') the main effect of treatment and the estimated propensity score, for example,
\begin{equation}\label{eq:psr}
E[Y|Z,X,\beta] = \beta_0 + \beta_{Z}Z + \beta_{e} e(X).
\end{equation}
If the treatment effect is modified by any covariates in $X$, then those interactions and an interaction between the treatment and the propensity score also must be included. The propensity score is typically estimated using logistic regression \citep{AlamSuperPS}, and taking $\widehat{e}(x)$ to be fitted values from that model. Unlike traditional outcome regression modelling, we need not assume that equation \eqref{eqn:outcomeReg} is correctly specified, but rather need that the propensity score model be correctly specified, and require also that any treatment effect heterogeneity is adequately captured by including interactions in $h(X, Z)$. Specifically, if
\[
h(X,Z) = \beta_0^\prime h_0(X) + \beta_1^\prime h_1(X,Z)
\]
then the treatment-free component $h_0(X)$ may be mis-specified, provided that the correct propensity score regression is constructed.  Suppose for example that the data generating model has conditional mean that contains an interaction with confounder $X_1$:
\begin{equation}\label{eq:int}
E[Y|Z,X,\beta] = \beta_0 + \beta_1 X_1 + \beta_{Z0}Z + \beta_{Z1} Z X_1.
\end{equation}
Then the propensity score regression model must contain terms that block the open backdoor paths that pass through the interactions; for example, this can be achieved using the model
\begin{equation}\label{eq:psrint}
E[Y|Z,X,\beta] = \beta_0 + \beta_{Z0}Z + \beta_{e0} e(X) + \beta_{Z1}Z X_1 + \beta_{e1} X_1 e(X).
\end{equation}
We expand upon this issue, and connections with semiparametric estimation, in Section \ref{sec:semipar}.

\subsubsection{Propensity score stratification or matching }
\label{sec:PSstrat}
While propensity score regression can be viewed as a means of parametric smoothing to make comparisons of individuals with different values of the treatment who share the same value of the propensity score, the non-parametric counterpart seeks to create strata of individuals with similar covariates and estimate the treatment effect within these strata, ultimately averaging the effect over the strata. If $X$ is very low-dimensional and composed of discrete variables, it may be feasible to create distinct strata for each combination of covariates without running into issues of data sparsity. However as the dimensionality of $X$ increases, this is not possible. The propensity score simplifies the analysis by providing a univariate summary of $X$ which can be used to stratify the sample. This may be done coarsely such as by using quintiles of the propensity score distribution \citep{RosenbaumRubin1984} or finely using, say, individually matched pairs \citep{Rubin1974}.

If matching rather than stratifying, additional considerations must be taken into account \citep{Stuart2010}. First, a matching criterion must be selected, such as the nearest neighbour or a random draw from within some `caliper' distance of the propensity score and a computational algorithm (greedy matching, optimal matching, etc.) must be selected. Matching need not be in pairs but could, for example, be 2:1 or 5:1. Matching is typically done without replacement, which prioritizes bias over efficiency.

Finally, an important note on implementation: matching is so often used for estimating the average treatment effect on the treated that this is the default procedure in many software packages such as \texttt{MatchIt} and \texttt{Matching} in \texttt{R}. Thus, when employing matching, one must take care to ensure the method used is appropriate to the chosen estimand.

\subsubsection{Inverse probability of treatment weighting} \label{sec:IPTW}
As noted in Section \ref{sebsec:Rubin}, inverse weighting was first developed to address missing data and then subsequently used to address confounding by viewing potential outcomes through a missing data lens. After estimating the propensity score, inverse probability of treatment weighting (IPTW) proceeds by fitting weighted averages (or a weighted regression) using weights constructed by $w_i=z\widehat{e}(x)^{-1} + (1-z)(1-\widehat{e}(x))^{-1}$. That is, each observation is weighted by the inverse probability of having received the observed treatment (not the ``$Z=1$" level of treatment). This is often described as constructing a ``pseudo-sample" in which there are no imbalances in the confounder distribution between treatment groups \citep{HernanEtAl2000}. IPTW can also be motivated by an importance sampling argument in which the target distribution is one in which $Z$ is independent of $X$ \citep{Saarela2015}. Similar arguments can be extended to estimate the ATT \citep{DR-AIPW-Stat}.

While IPTW relies on correct specification of the propensity score only, it may be very inefficient relative to a correctly specified outcome regression model. The \textit{augmented} IPTW estimator reduces variability in the IPTW estimator while simultaneously providing additional safeguards against model misspecification \citep{Scharfstein1999,BangRobins2005}. An augmented estimator can also be computed by fitting an unweighted regression model that includes the difference in inverse probability of treatments, i.e.~$z\widehat{e}(x)^{-1} - (1-z)(1-\widehat{e}(x))^{-1}$, as a covariate. See Section \ref{sec:dr} for more on \textit{double robustness}.

\subsubsection{Model checks}
All of the methods described above rely on identifiability assumptions which include no unmeasured confounding and positivity. They further make assumptions about correct model specification, whether that is of the outcome model or the propensity score. When the estimation approach relies on correct specification of the outcome model, all standard residual diagnostics should be employed, although it can be difficult to detect some forms of misspecification and positivity cannot be reliably assessed \citep{KangSchafer2007}.

When using propensity score methods, checking balance between treatment groups can be achieved by visually comparing distributions of covariates \citep[e.g.][]{Tan2006} or using distributional summaries such as the standardized mean difference \citep{CochraneChap6:smd}. Positivity can be assessed by visual examination of the propensity score distribution, checking whether the distributions among the different treatment groups are overlapping. For a worked example using all of the above-described methods including illustration of the model checks, see the tutorial of \citet{STRATOSpaper}. There is no straightforward way to assess positivity when $X$ is even moderately high-dimensional without relying on a propensity score. Thus, while traditional statistical methods of analysis such as regression can be employed for estimation, use of the propensity score still proves useful.

\subsection{Double robustness}\label{sec:dr}

The estimation procedures described above typically depend on the specification of two models for the observable quantities; the outcome model, representing the conditional distribution of $Y$ given $(X,Z)$, and the treatment model, representing the conditional distribution of $Z$ given $X$.  Doubly robust procedures provide consistent estimation of the ATE if at least one of these models is correctly specified and match the data generating structure.  The augmentation strategy from Section \ref{sec:IPTW} is an example of a doubly robust procedure, as the resulting estimator is consistent provided at least one of the augmenting model or the propensity score-based re-weighting model is correctly specified.  Similarly, elaborations of the propensity score regression model in \eqref{eq:psr} can also be considered doubly robust; if the outcome model
\begin{equation}\label{eq:psrdr}
E[Y|Z,X,\beta] = \beta_0 + \beta_{Z}Z + \beta_{e} e(X) + \mu_0(X,\beta_X)
\end{equation}
is used, and in reality the data generating model has outcome with conditional mean $\beta_{Z}Z + \mu_0(X,\beta_X)$, then the least squares estimator for $\beta_Z$ derived from \eqref{eq:psrdr} is also consistent even if the model for $e(x)$ is mis-specified.

The lure of doubly robust procedures is evident.  However, we must remember that any allusion to `correct specification of the outcome model' (one of the aforementioned `robustnesses') is in conflict with the principal reason that propensity score adjustment is deployed; if we acknowledge the possibility that the outcome model is correctly specified, then by standard theory, the optimal inference approach is the one that uses this outcome model without any further elaboration.

\section{Causal inference under semiparametric specifications} \label{sec:semipar}

Estimation using the models referred to in Section \ref{sec:estimate} is typically carried in a semiparametric inference setting, without distributional assumptions about outcome residual errors or confounders.  For the linear regression models such as model \eqref{eq:psr}, estimation is typically performed via least squares.  Consequently, the optimality of inference approaches in terms of efficiency is an important area of study; the form of optimal (if not always feasible) estimators of causal parameters can be deduced using semiparametric efficiency theory; see for example, \cite{BKRW}.   This is now much studied and relied upon in the causal literature, and has close connections with techniques with a rich history that were developed as extensions of parametric inference methods such as estimating functions in general \citep{godambe1991estimating} and generalized estimating equations \citep{LiangZeger1986} in particular, as well as parallel developments in econometrics (see \cite{newey90} for a summary).

The model of \eqref{eq:psr} provides a simple setting in which the semiparametric results can be illustrated.  Suppose the causal (structural) model is
\begin{equation}\label{eq:RMN0}
Y_i = \beta_Z Z_i + h_0(X) + \epsilon_i
\end{equation}
where $h_0(X)$ is some function of the confounders which is unknown to the analyst, and $\epsilon_i$ satisfies $E[\epsilon_i|X_i] = 0$ and $Var[\epsilon_i |X_i] = \sigma^2(X_i) < \infty$.  The key paper of \cite{Robins1992} shows that the estimating equation
\begin{equation}\label{eq:RMN0ee}
\sum_{i=1}^n (z_i - e(x_i))(y_i - \beta_Z z_i) = 0
\end{equation}
with solution
\[
\widehat \beta_Z = \frac{\sum\limits_{i=1}^n (z-e(x_i)) y_i}{\sum\limits_{i=1}^n (z-e(x_i)) z_i}
\]
results in consistent inference for $\beta_Z$; see also \cite{newey90}.   The estimating equation in \eqref{eq:RMN0ee} does not attempt to model the treatment-free component $h_0(X)$, but by considering the (inference) regression model
\begin{equation}\label{eq:RMN1}
Y_i = \beta_Z Z_i + \mu_0(X_i,\phi) + \epsilon_i
\end{equation}
the estimating equation can be modified to
\begin{equation}\label{eq:RMN1ee}
\sum_{i=1}^n (z_i - e(x_i))(y_i - \beta_Z z_i - \mu_0(x_i,\widehat \phi)) = 0
\end{equation}
yielding the solution
\[
\widehat \beta_Z = \frac{\sum\limits_{i=1}^n (z-e(x_i)) (y_i-\mu_0(x_i,\widehat \phi))}{\sum\limits_{i=1}^n (z-e(x_i)) z_i}
\]
which corresponds to an estimator that typically will have a variance no larger than that of the original estimator.

There is a connection between this approach and the propensity score regression model in \eqref{eq:psr}.  Suppose for simplicity the intercept in model \eqref{eq:psr} is omitted, and the parameters estimated using least squares.  The corresponding estimating system becomes
\[
\sum_{i=1}^n \begin{pmatrix}
  z_i \\
  e(x_i)
\end{pmatrix} (y_i -  \beta_Z z_i - \beta_e e(x_i)) = \mathbf{0}.
\]
Subtracting the second equation from the first yields the equivalent form
\[
\sum_{i=1}^n \begin{pmatrix}
  z_i-e(x_i) \\
  e(x_i)
\end{pmatrix} (y_i - \beta_Z z_i - \beta_e e(x_i)) = \mathbf{0}.
\]
Thus, choosing $\mu_0(x,\phi) = \phi e(x)$ in \eqref{eq:RMN1} corresponds to the model in \eqref{eq:psr}.  Extension to more complicated linear specifications, including those that involve interactions between treatment and confounders, is straightforward.  We may conclude that the propensity score regression approach is a version of the semiparametric formulation.

Equation \eqref{eq:RMN0ee} is based on the $m$-estimating function
\[
m(X,Y,Z;\beta_Z) = (Z-E[Z|X])(Y - \beta_Z Z)
\]
and it is evident that, provided the term $(Y - \beta_Z Z)$ is independent of $Z$ (that is, specifically,  that the effect of treatment is correctly captured in \eqref{eq:RMN0}) then $E[m(X,Y,Z;\beta_Z)] = 0$ if the model $E[Z|X] = e(X)$ is correctly specified; in these calculations, expectations are taken with respect to the structural (data generating) model.

As summarized by \cite{newey90} \cite[see also][chap. 3]{Tsiatis_Book} the semiparametric efficient estimator of $\beta_Z$ is obtained by considering the influence function
\[
\varphi(X,Y,Z;\beta_Z) = -\frac{1}{E[\dot{m}(X,Y,Z;\beta_Z)]} m(X,Y,Z;\beta_Z)
\]
where here
\[
\dot{m}(x,y,z;\beta_Z) = \frac{\partial m(x,y,z;\beta_Z)}{\partial \beta_Z} = - z(z - e(x))
\]
and then solving the estimating equation
\[
\sum_{i=1}^n \varphi(x_i,y_i,z_i;\beta_Z) = 0
\]
which, in this case, matches \eqref{eq:RMN0ee} as $\dot{m}(x,y,z;\beta_Z)$ does not depend on $\beta_Z$. Thus, under standard conditions, the estimator that results from \eqref{eq:RMN0ee} is consistent and asymptotically Normally distributed with asymptotic variance
\[
\frac{E[\{m(X,Y,Z;\beta_Z)\}^2]}{\{E[\dot{m}(X,Y,Z;\beta_Z)]\}^2} = \frac{E[(Z-E[Z|X])^2(Y - \beta_Z Z)^2]}{E[Z(Z-e(X))]^2} =
\frac{E[Var[Z|X] v(X)]}{\{E[Var[Z|X]]\}^2}
\]
where $v(X) = E[(Y-Z \beta_Z)^2|X] = \{h_0(X)\}^2 + \sigma^2(X)$. In this expression, $Var[Z|X] = e(X) (1-e(X))$, and it is evident that if $\sigma^2(X) = \sigma^2$ and we have homoscedastic residual error structure, then
\[
\frac{E[Var[Z|X] v(X)]}{\{E[Var[Z|X]]\}^2} \geq \frac{\sigma^2}{E[e(X)(1-e(X))]}
\]
which provides a lower bound on the variance of the propensity score regression estimator.

A similar semiparametric efficiency analysis of the IPTW estimator from Section \ref{sec:IPTW} can be carried out; see for example \citet{HIR2003}, who provide the asymptotic variance for the IPTW estimator of the ATE $\beta_Z$ in the model \eqref{eq:RMN0} as
\[
E \left[\frac{\sigma_0^2(X)}{1-e(X)}  + \frac{\sigma_1^2(X)}{e(X)}\right]
\]
where $\sigma_j^2(X)$ is the residual variance for $Z=j, j=0,1$, which under homoscedasticity reduces to
\[
\sigma^2 E \left[\frac{1}{e(X)(1-e(X))}\right].
\]
There are several things to note about these results.  First, the lower bound for the regression estimator is smaller than the lower bound for the IPTW estimator and if in fact $h_0(X) \equiv 0$, the lower bound is achieved. This result was noted by \citet{Ertefaie2011}. Secondly, the IPTW method does not rely upon a need for correct specification of the treatment effect term.  Finally, these results for both methods can be extended to procedures involving augmentation, and also to cases where the treatment effect model reflects modification by (that is, interaction with) confounders that include additional treatment effect parameters.

If the propensity score model is represented by a parametric form, say $e(x) \equiv e(x;\gamma)$, then in the estimation we may replace $e(x)$ by $\widehat{e}(x) \equiv e(x;\widehat \gamma)$ where $\widehat \gamma$ is obtained from an estimating procedure that depends on the $x$ and $z$ data only,  and proceed with the `feasible' estimation that solves the corresponding version of \eqref{eq:RMN0ee}.  As shown by \cite{HenmiEguchi2004}, the use of this plug-in estimator does not affect the asymptotic variance of the estimator as $\widehat \beta_Z$ and $\widehat \gamma$ are asymptotically independent.

\section{Modern developments and current challenges}
In this section, we point to some current directions for research in causal inference. In some instances, these directions involve weakening or violations of the identifiability assumptions listed above. In others, they rely on the traditional identifiability assumptions and require additional similar assumptions to further account for new data structures that have arisen in step with increase digital storage capacity or computational power.

\subsection{Unmeasured confounding}

The presence of unmeasured confounding disrupts causal, and indeed most conventional, statistical analyses.  Unmeasured confounders are the `unknown unknowns' of statistics; we don't know if they exist, and furthermore we cannot always be sure that we will ponder their existence.  Figure \ref{fig:UMCs}, left panel, demonstrates the simplest unmeasured confounding setting; even if $X$ is conditioned upon, the open path from $Z$ to $Y$ via $U$ allows confounding of the treatment effect.  In a standard analysis, there is no way of overcoming the unmeasured confounding as data on $U$ are not available; it is possible to examine the effect of a possible hypothetical unmeasured confounder using simulation methods in sensitivity analysis.  An alternative approach can be derived if there exists a further (observable) variable, $W$, that (a) is a cause of $Z$, (b) is independent of $U$, and (c) has no direct effect on $Y$; see the right panel of Figure \ref{fig:UMCs}.

\tikzset{
    -Latex,auto,node distance =1 cm and 1 cm,semithick,
    state/.style ={draw, minimum width = 0.7 cm},
    box/.style ={rectangle, draw, minimum width = 0.7 cm, fill=lightgray},
    point/.style = {circle, draw, inner sep=0.08cm,fill,node contents={}},
    bidirected/.style={Latex-Latex,dashed},
    el/.style = {inner sep=3pt, align=left, sloped}
}
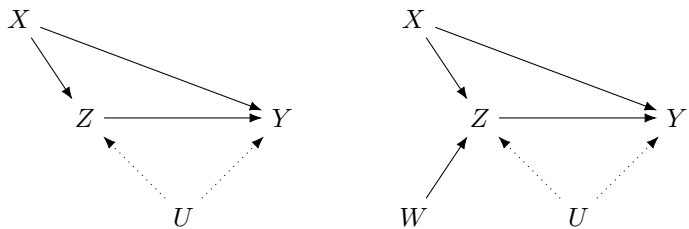
\begin{figure}[h]
\centering
\begin{tikzpicture}[scale=1.75]
    \node (x1) at (-1.5,1.75) {${X}$};
    \node (y1) at (0.5,1){${Y}$};
    \node (z1) at (-1,1) {${Z}$};
    \node (u1) at (-0.25,.25) {${U}$};

    \path (x1) edge (z1);
    \path (x1) edge (y1);
    \path[dotted] (u1) edge (z1);
    \path[dotted] (u1) edge (y1);
    \path (z1) edge (y1);

    \node (x2) at (1.5,1.75) {${X}$};
    \node (y2) at (3.5,1){${Y}$};
    \node (z2) at (2,1) {${Z}$};
    \node (u2) at (2.75,.25) {${U}$};
    \node (w2) at (1.5,.25) {${W}$};

    \path (x2) edge (z2);
    \path (x2) edge (y2);
    \path[dotted] (u2) edge (z2);
    \path[dotted] (u2) edge (y2);
    \path (z2) edge (y2);
    \path (w2) edge (z2);

 \end{tikzpicture}
    \caption{Unmeasured confounding.  Left: there is an unmodellable open path from $Z$ to $Y$ via the unmeasured confounder $U$. Right: instrumental variable $W$ that is a predictor of $Z$ but is independent of $U$ and has no direct effect on $Y$. }
    \label{fig:UMCs}
\end{figure}

For example, consider a randomized experimental study of a binary treatment with imperfect compliance; in such a study, there are no measured confounders.  In the study, where study participants are randomly assigned to one of two treatment groups, each participant may adhere to the treatment assignment or contravene the treatment assignment.  If $W$ is the treatment assignment indicator, and $Z$ records the treatment actually taken, then $W$ is an instrument, and even if adherence to assigned treatment is determined by unmeasured factor $U$, $W$ is independent of $U$ by design.  Of course, the intention-to-treat analysis that compares outcomes for the two assigned treatments can still be carried out, however it may not necessarily reflect the target of inference.

Such a variable $W$ is called an instrumental variable.  This variable can be deployed in statistical analysis, using a number of different techniques, to recover consistent estimators of the causal effect of $Z$ on $Y$ if the required conditions hold.   Instrumental variable methods are sometimes claimed to be able to `overcome' the issue of unmeasured confounding.  Perhaps a more honest view is that instrumental variable methods solve the causal inference problem by relying on a different set of equally strong assumptions from those listed in Section \ref{sec:Assess}. In particular, it is only under strong assumptions that an instrumental variable analysis can be used to estimate an ATE or ATT. Otherwise, the analysis targets the `complier average causal effect' (CACE), which is the effect of treatment among those individuals whose observed treatment agrees with their assigned (by the instrument) treatment -- a latent subgroup that therefore complicates interpretation.

There are two connections with the earlier described methods that are noteworthy.  First, the ATT quantity from Section \ref{sebsec:msm} is an estimand which may be estimated using instrumental variable methods; see \cite{ImbensAngrist1994-LATE, AngristEtAl1996}.  Secondly, the propensity score regression estimator derived from \eqref{eq:RMN0ee} can also be justified as an instrumental variable estimator using the instrument $W = Z - e(X)$.

\subsection{Interference}
Interference occurs when an individual's potential outcome depends on not only their treatment assignment, but that of others'. The presence of interference is a clear violation of SUTVA, and requires an extended potential outcome notation and specialized methods. Currently, estimation has focused on methods of simplifying the way in which interference can occur, such that each individual's treatment can be viewed as a bivariate vector whose components consist of the treatment received directly by that individual and that received indirectly via spillover, where the latter could represent a proportion or the number of treated `others', often termed neighbours, who affect that individual. Estimands then focus on the impact of direct treatment assuming a fixed level of interference (e.g., the average outcome if treated versus not, given 30\% of one's  neighbours are treated), as well as the indirect impact of treatment (e.g., the average outcome if untreated and, say, 50\% of one's  neighbours are treated as compared to if untreated and no neighbours are treated).

Hudgens and Halloran (\citeyear{hudgens_toward_2008}; \citeyear{halloran_causal_2012,halloran_dependent_2016}; \citeyear{saul_chapter_2017}) have developed many critical innovations in this realm, including the notion of \textit{partial} interference, which supposes that the data are composed of many (often small) fully connected clusters of individuals such that spillover occurs within but not between clusters. Alternatively, some authors have relaxed SUTVA \citep{vanderLaanNetwork2014, Forastiere2021}, and instead rely on the Stable Unit Treatment on Neighbourhood Value Assumption (SUTNVA). SUTNVA assumes that, within a social network, interference or spillover of a treatment can occur only from an immediate connection, termed an individual's `neighbourhood'. This literature requires knowledge of the network of connections, which could be defined by, for example, self-reported friendships or social media connections.

\subsection{Sequential Multiple Assignment Randomized Trials}
An area of growing interest is that of algorithmic decision-making and decision-support systems. While these may be ubiquitous in many online activities, the use of individually-tailored decisions in other fields such as the health sector call for a high level of care and scrutiny. Within the statistical literature, this area is often termed precision medicine, and methodologies have centred on estimation and inference for \textit{dynamic treatment regimes} or adaptive treatment strategies. Estimation of optimal treatment strategies, particularly for sequences of treatments in which there may be delayed effects and interactions between treatments taken at different points in time, has grown predominantly from the causal inference literature, relying on non-experimental data with challenges such as time-dependent confounding and mediation. However, there remains an important role for randomized trials, using Sequential Multiple Assignment Randomized Trials (SMARTs) \citep{CollinsEtAl2007,OettingEtAl2011,LeiEtAl2012,Kidwell2014} to study sequences of treatments tailored to individual patient characteristics.

The number of SMARTs that have been conducted is increasing, though the design is still not common and further innovations are needed. SMARTs are often used to determine optimal medication sequences or educational interventions which may best be delivered via a cluster randomized trial -- e.g.~enlisting physicians who will recruit patients from their clinic in the first case or randomizing educators (or schools) in the second. While a very small number of cluster randomized SMARTs have been carried out \citep{Kilbourne2014}, they are rare and their design operating characteristics remain not fully explored. Similarly, adding adaptive components to the trial design may be of interest to allow for early stopping of a SMART \citep{Cheung2015}.

\subsection{Electronic health records and other not-for-research data sources}
Increasingly, data are collected and stored from a wide variety of sources. For example, many individuals voluntarily share data by using or uploading information to fitness tracking apps, or through loyalty programs at retailers. While ever increasing storage and computational capacity has created opportunities for new analyses, 
the increased availability and quantity of data does not necessarily correspond with increased quality of data, and so causal thinking -- and analyses aimed at reducing or eliminating biases -- are required.

Consider electronic health records (EHRs), in which health data are recorded on an often large population, logging information on interactions with a healthcare provider, laboratory and other physiologic measurements, and treatments prescribed \citep{EHRs}. These data may be subject to biases due to a variety of causes including lack of generalizability, covariate-driven patterns of observation (or missingness), measurement error and misclassification, and confounding. Further, statistical methods to ensure data security and privacy may need to be developed if EHRs from multiple sites are combined.

Generalizability of EHR data will typically depend on the source of the data. EHRs drawn from health registries in countries or regions with national or single-payer healthcare provision (e.g., the National Health Service in the United Kingdom, provincial healthcare in Canada) are more likely to be representative of the population under study than registries of privately or employer-insured populations, or a population receiving subsidized medical care (e.g.,~Medicaid in the United States).

A common feature of EHRs, particularly those observed in a diverse population, is that observations may be irregular. Some patients may be observed frequently, due to a high number of comorbid conditions or medications that require close monitoring. Other patients may be observed less frequently, e.g.~individuals with inflexible working hours. These covariate-driven monitoring patterns can lead to certain patient groups being over-represented in the data which, if not addressed in the analysis, can bias estimators \citep{Coulombe2021Biom,Coulombe2022AAS}.

In terms of measurement error or misclassification, 
EHRs may be subject to misclassification of medication usage as this is not often available in these records. Rather, EHRs typically record drugs \textit{prescribed}, but may not record drugs \textit{dispensed} (pharmacy claims). Where both prescription and acquisition of the treatment are available, actual usage is rarely known, and must be inferred from algorithms that make assumptions regarding overlapping prescriptions 
and gaps (`grace' periods) between prescriptions. For example, two treatments may be regarded as consecutive or maintained if no more than four weeks pass between the end of one prescription and the beginning of another and otherwise be deemed separate treatment episodes \citep{EHR-Rx}.


\subsection{The role of machine learning in causal inference}

Machine learning methods are revolutionizing many application areas, and are at the centre of a huge research enterprise encompassing aspects of statistics, mathematics, and computer science.  Statisticians play a vital role in this enterprise, particularly in developing frameworks for understanding the operating characteristics of machine learning methods when applied to data.  Many aspects of the most successful machine learning approaches have their foundations in the statistical literature (for example, hierarchical models, Gaussian processes, stochastic approximation methods, Dirichlet processes) but their current applicability, at such scale and in such complexity, require sophisticated algorithmic strategies, and also developments in theoretical understanding that have originated in the machine learning literature.

The impact of machine learning, and artificial intelligence in general, in health applications for example is already widely felt in public and private sectors, and in fact has a long history, stemming at least from developments in the probabilistic and computational treatment of graphical models and Bayesian networks originating more than forty years ago (see, for example, \cite{SHORTLIFFE1975351}).   There are, however, limits to what can be achieved by the `automatic' application of mathematical, statistical, or computational approaches alone.  In the current context, it is not wholly unfair to characterize many modern machine learning methods as prediction machines that can produce superlative reconstructions of observed data, as well as considerable ability to `generalize' those predictions out-of-sample.  Whereas this has evident attraction in many domains, it is also that evident prediction is not at the heart of causal reasoning.

From one perspective, the use of machine learning methods as flexible prediction tools can be helpful in constructing the models described in Section \ref{sec:estimate} that can be then deployed to estimate treatment effects; see \citet{KennedyChapter2016} for a clear review of the empirical process theory that governs the behaviour of nonparametric estimators of nuisance functions in causal inference.  However, this needs to be done with care, as the statistical properties of the flexible fitting procedure are not always straightforward to understand.

\textit{Where} to deploy machine learning within the causal framework also requires careful thought. Better estimation, or prediction, of the outcome model can lead to superior performance, with reduced variability and/or bias, in doubly robust or outcome-modelling based methods (see, for example, \citet{AtheyImbens2016,hahn2020bayesian}). However, uncertainty representation (and consequently testing) based on machine learning-based fitting needs very careful calibration; kernel-based methodologies \citep{daili2021} offer a way forward as they are flexible yet statistically tractable. Further, using machine learning methods to construct the treatment assignment/propensity score model requires particular care and is arguably ill-advised, since the primary objective of propensity score modelling is to create balance. A propensity score that predicts treatment with high precision may lead to a lack of overlap and hence positivity violations \citep{AlamSuperPS}.  Also, as in other domains, there are ethical aspects to consider when using machine learning-based analysis \citep{ShortreedMoodie-StatSci}.  For example,  the use of algorithm-generated treatment decisions that are not transparent in their construction may lead to the perpetuation of errors made in the training data.

\section{Concluding remarks}\label{Discussion}

As we have elaborated above, causal inference in statistics has formalized, mathematized, and operationalized notions of `fairness' so that causal conclusions can be supposed from analyses using imperfect data that are subject to confounding, missing data or selection biases, measurement error and more. Causal inference should not be seen as conflicting with traditional statistical methods of explanatory analyses, and causal thinking should be considered relevant even in predictive modelling or machine learning applications for it is not only the underlying data generating structure that leads to potential biases such as confounding, but also the data collection and research design that can give rise to inappropriate inferences and decision-making if not appropriately taken into account.

\section*{ACKNOWLEDGEMENTS}
EEMM and DAS acknowledge funding from Discovery Grants from the Natural Sciences and Engineering Research Council of Canada. EEMM is supported by a career award from the Fonds de recherche du Qu\'ebec - Sant\'e and holds a Canada Research Chair in Statistical Methods for Precision Medicine.

\bibliographystyle{chicago}
\bibliography{All-References-Causal2022}

\end{document}